\documentclass[12pt]{article}

\usepackage[a4paper,left=30mm,right=20mm,top=20mm,bottom=20mm]{geometry}
\usepackage{graphics}
\usepackage{amsmath}
\usepackage{epsfig}
\usepackage{cite}
\usepackage{lineno}
\newcommand{\be}{\begin{eqnarray}}
\newcommand{\ee}{\end{eqnarray}}

\title{Probing proton intrinsic charm in photon or $Z$ boson production accompanied by heavy jets at LHC}
\author{A.V.~Lipatov$^{1,2}$, G.I.~Lykasov$^{2}$, Yu.Yu.~Stepanenko$^{2}$, V.A.~Bednyakov$^{2}$}

\begin{document}

\maketitle

\begin{center}

{\it $^1$Skobeltsyn Institute of Nuclear Physics, Moscow State University, 119991 Moscow, Russia}\\
{\it $^2$Joint Institute for Nuclear Research, Dubna 141980, Moscow region, Russia}\\
\end{center}

\vspace{0.5cm}

\begin{center}

{\bf Abstract }

\end{center} 

\indent

We consider an observable very sensitive to the non-zero intrinsic charm (IC) contribution to the proton density.
It is the ratio between the differential cross sections of the photon or $Z$-boson and $c$-jet 
production in the $pp$ collision, $\gamma(Z) + c$, and the $\gamma(Z)$ and the $b$-jet production.
It is shown that this ratio can be approximately flat or increasing at large $\gamma(Z)$ transverse momenta $p_T$ 
and their pseudo-rapidities $1.5 < \eta < 2.4$ if the IC contribution is taken into account.  
On the contrary, in the absence of the IC this ratio decreases as $p_T$ grows.    
We also present the ratios of the cross sections integrated over $p_T$ as a function of the IC probability $w$. 
It is shown that these ratios are mostly independent on the theoretical uncertainties, and such predictions could 
therefore be much more promising for the search for the intrinsic charm signal at the LHC compared to the predictions 
for $p_T$-spectra, which significantly depend on these uncertainties.

\vspace{1.0cm}

\noindent
PACS number(s): 12.15.Ji, 12.38.Bx, 13.85.Qk

\newpage
\indent

\section{Introduction} \indent

The hypothesis of the {\it intrinsic} (or valence-like) heavy quark component, the
quark Fock state $|uudQ{\bar Q}\rangle$\cite{1,2,3,4} in a proton  
suggested by Brodsky with coauthors\cite{1,2} (BHPS model) is intensively discussed in 
connection with an opportunity to verify it experimentally\cite{5,6,7,8,9,10,11,12,13,NNPDF,14}. 
Up to now, there is a long-standing debate about the possible existence 
of the {\it intrinsic} charm (IC) and {\it intrinsic} strange (IS) 
quarks in a proton\cite{7,14,15,16}.      
Thorough theoretical and experimental studies of these 
intrinsic heavy quark components 
would be very important for the experiments performed at the LHC.  
    
Recently it was shown that the possible existence of the intrinsic heavy quark 
components in the proton can be seen not only in the inclusive heavy flavor
production at high energies\cite{8}, but also 
in the semi-inclusive production of prompt photons or vector bosons 
accompanied by heavy quark jets\cite{9,13}.
An experimental hint on possible existence of the IC contribution was observed
in the Tevatron experiment on the prompt photon production in the association of the $c$
and $b$ jets in the $p{\bar p}$ annihilation at $\sqrt{s} = 1.98$~TeV\cite{17,18}. It was shown
that the description of the Tevatron data within the  perturbative QCD (pQCD)  
could be significantly improved if the IC contributions were taken into account
The photon transverse momentum ($p_T$) spectrum in the $\gamma+c$ production and the ratio
of the spectra in the $\gamma+c$ and $\gamma + b$ production measured at the Tevatron\cite{20} 
are better described within the BHPS model\cite{1,2}, which includes the IC 
contributions. According to the pQCD calculations\cite{19}, in the absence of the IC contribution this ratio
decreases, when $p_T$ grows, while the Tevatron data show its flat behavior at large 
$p_T \ge 100$~GeV\cite{20}.     

The possible IC signal can also be observed in the hard $pp$ production of the gauge bosons $Z$ or $W$ 
accompanied by heavy flavors. As it was shown\cite{13}, the ratio of the $Z + c$ and $W$ + heavy jet
production cross sections maximizes the sensitivity to the IC 
component of the proton. Our early predictions about a possible intrinsic charm signal in the production  
of prompt photons or gauge bosons accompanied by heavy flavor jets 
concerned their transverse momenta distributions in the mid-rapidity 
region of $pp$ collisions at the LHC energies\cite{9,13}.
It was obtained with the IC probability about $w = 3.5$\%, 
which is the upper limit being due to constraints from the 
HERA data on the deep inelastic scattering. 
However, the upper limit of the IC probability in a proton is still
very actively debated\cite{7,14,15,16}. 
Therefore, in the present paper we focus mainly on the predictions for searching at any $w$ for the IC signal
in the observables, which are very little sensitive to the theoretical uncertainties, namely, the ratios
between the $\gamma(Z) + c$ and $\gamma(Z) + b$ cross sections in $pp$ collisions at the LHC energies.
An important advantage of these observables is that many theoretical uncertainties, for example, heavy quark
masses, the factorization and/or renormalization scales, are canceled, as will be demonstrated below.
We show that the measure of these ratios is much more promising for the search for the IC signal.  
      
Below we perform the calculations in two ways.
First, we use the parton-level Monte Carlo event 
generator \textsc{mcfm}\cite{21}, which implements the 
NLO pQCD calculations of associated $Z$ boson and heavy flavor
jet production. The detailed description of 
the \textsc{mcfm} routine is available\cite{21}.
To generate the prompt photon and heavy jet production cross sections,
we apply the $k_T$-factorization approach\cite{22,23},
which becomes a commonly recognized tool in the high energy
phenomenology.
Our main motivation is that it gives a better description of the 
Tevatron data compared to the NLO pQCD calculations\cite{19},
as it was claimed\cite{17,18}.
We apply this approach to the associated $Z$ and heavy 
jet production to perform an independent cross-check of our 
results\footnote{Unfortunately, the \textsc{mcfm} routine 
does not produce the prompt photon and heavy jet production cross sections.}.

The outline of our paper is the following. In Sections~2 and 3 we recall  
basic ideas with a brief review of calculation steps. In Section~4 we present
the numerical results of our calculations and a discussion. Finally, Section~5
contains our conclusions.

\section{Intrinsic charm density in a proton as a function of IC probability $w$} \indent

According to\cite{6,12,24}, the intrinsic charm distribution at 
the starting scale $\mu_0^2$ as a function of $x$ can be presented in the following 
approximated form:
\begin{eqnarray}
c_{int}(x,\mu_0^2) = c_0 w x^2 \left[(1-x)(1+10x+x^2)+6x(1+x)\ln(x) \right],
\label{def:fcPumpl}
\end{eqnarray} 

\noindent 
where $w$ is the probability to find the Fock state $|uudc{\bar c}\rangle$ in the proton, 
$c_0$ is the normalization constant and the masses of the light quarks and the nucleon are neglectedable 
compared to the charm quark mass.   
The inclusion of the non-zero nucleon mass leads to a more complicated analytic form\cite{25}.
According to the BHPS model\cite{1,2},
the charm density in a proton is the sum of the {\it extrinsic} and {\it intrinsic} charm densities, 
\begin{eqnarray}                              
xc(x,\mu_0^2) = xc_{\it ext}(x,\mu_0^2) + xc_{\it int}(x,\mu_0^2).
\label{def:cdens_start}
\end{eqnarray}

\noindent
The {\it extrinsic}, or ordinary quarks and gluons are generated on a short-time scale associated
with the large-transverse-momentum processes. Their distribution functions 
satisfy the standard QCD evolution equations.
Contrariwise, the {\it intrinsic} quarks and gluons  
can be associated with a bound-state hadron dynamics and one believes that they 
have a non-perturbative origin. The lifetime of this Fock-state should be much more 
than the interaction time of the hard probe\cite{25}. Some comments on this are presented
below.
It was argued\cite{2} that existence, for example, of intrinsic 
heavy quark pairs $c{\bar c}$ and $b{\bar b}$ within the 
proton state can be due to the gluon-exchange and 
vacuum-polarization graphs. 

\begin{figure}
\begin{center}
\epsfig{figure=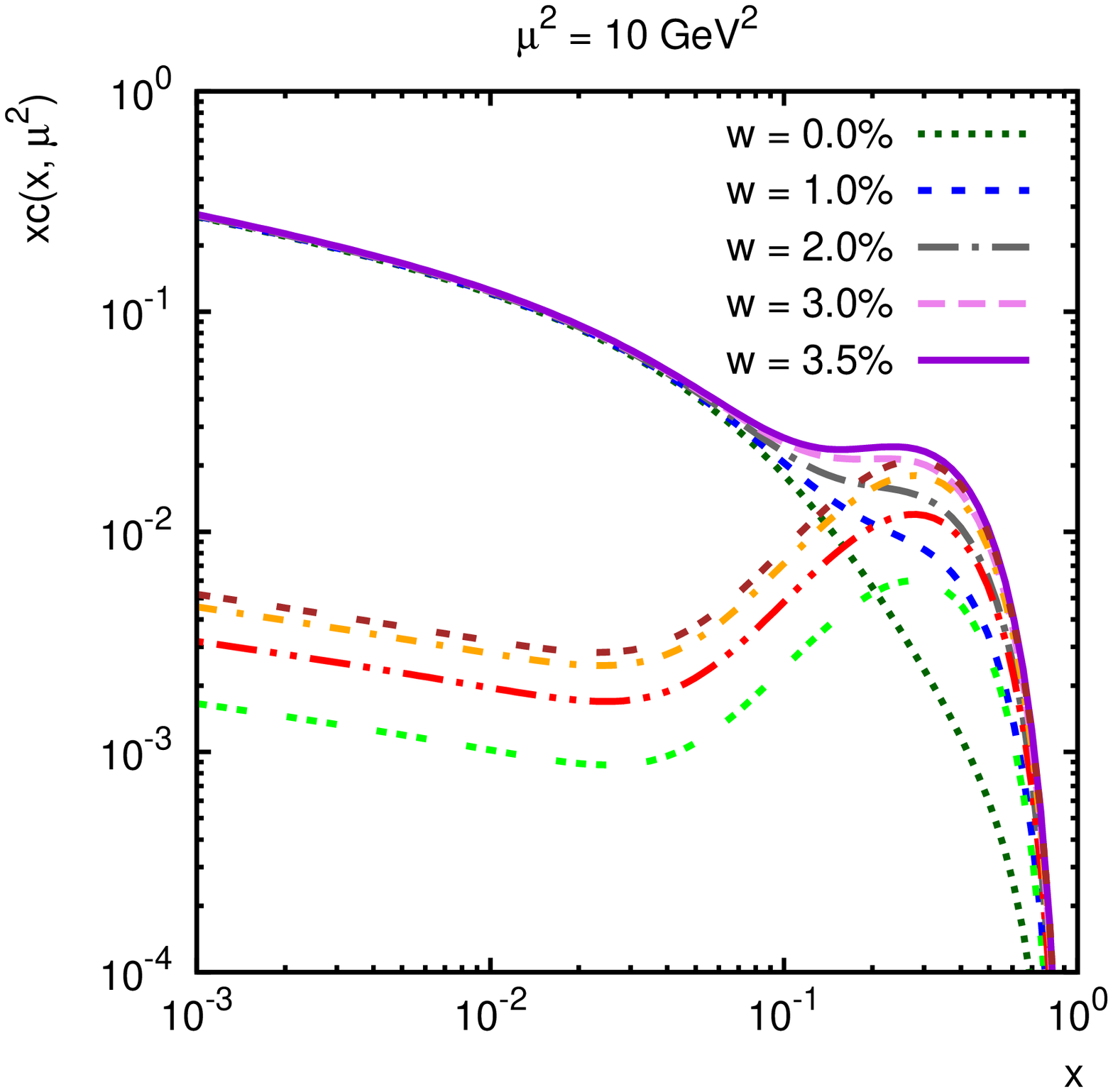,width = 10cm, angle = 270}
\hspace{-1.0cm}
\epsfig{figure=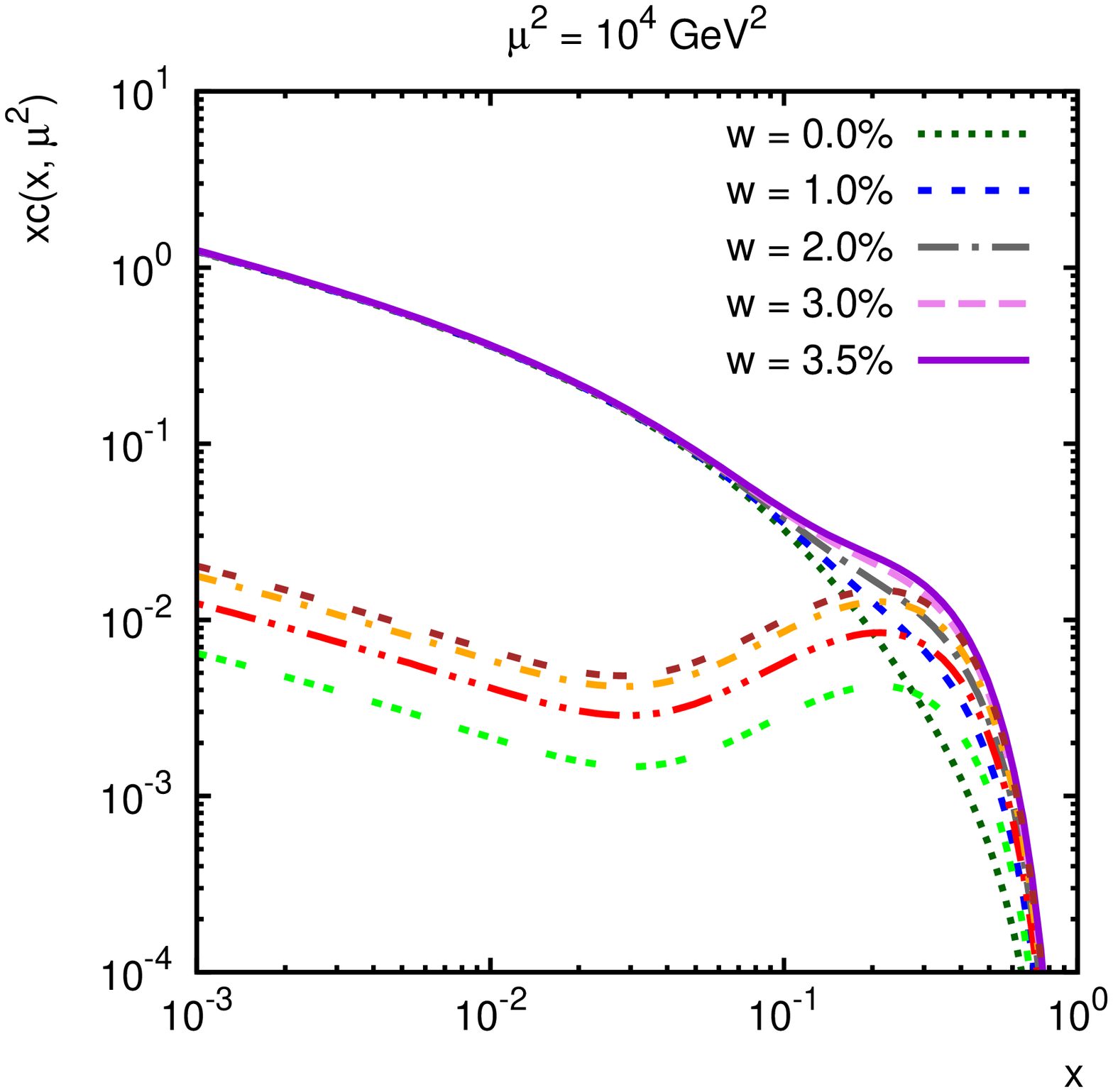,width = 10cm,angle = 270}
\caption{ 
The total charmed quark densitiy $xc(x,\mu^2)$ as a function of $x$ at different values of $w$
at $\mu^2 = 10$~GeV$^2$ (top) and $\mu^2 = 10^4$~GeV$^2$ (bottom).
The triple-dashed line is the IC contriubion at $w=$ 1\%, the dashed-double-dotted line corresponds to
the IC at $w=$ 2\%, the dashed-dotted curve is the IC at $w=$ 3\% and the double-dashed line corresponds to
the IC at $w=$ 3.5 \%.
}
\label{fig1}
\end{center}
\end{figure}

The charm density $xc(x,\mu^2)$ at an arbitrary scale $\mu^2$ is calculated 
using the Dokshitzer-Gribov-Lipatov-Altarelli-Parisi (DGLAP) equations\cite{26}.
Let us stress here that both the intrinsic part $xc_{\it int}$ and extrinsic one $xc_{\it ext}$ 
depend on $\mu^2$.
In the general case, there is some mixing between two parts of (\ref{def:cdens_start}) during 
the DGLAP evolution. However, such mixing is negligible\cite{12,IC_evol}, especially at large 
$\mu^2$ and $x$.
It can be seen from comparison of our calculations of charmed quark densities presented 
in Fig.~\ref{fig1}, where this mixing was included within the CTEQ\cite{27} set, and 
Fig.~2 of\cite{12}, when the mixing between two parts of the charm density was neglected.
Our results on the total charm density $xc(x,\mu^2)$ are in good agreement 
with the calculations\cite{12} in the whole kinematical region of $x$ because at $x<0.1$ the IC contribution
is much smaller than the extrinsic one.   
Therefore, one can apply the DGLAP evolution separately to the first part $xc_{\it ext}(x,\mu_0^2)$
and the second part $xc_{int}(x,\mu_0^2)$ of~(\ref{def:cdens_start}), 
as it was done in \cite{12,IC_evol}.
Such calculations were done by the CTEQ\cite{27} and CT14\cite{28} groups
at some fixed values of the IC probability $w$. Namely,
the CTEQ group used $w = 1$\% and $w = 3.5$\%, and CT14 used $w = 1$\% and $w = 2$\%.

Note that, according to the recent paper\cite{25}, the lifetime of the intrinsic charm should 
be more than the interaction time, at least, by a factor of about $5$, when the quark Fock-state can be
observed with the satisfactory accuracy.
The ratio of these times is proportional to $Q^2$ or $p_T^2$\cite{25}.     
We will analyze the hard processes of $\gamma(Z)$ production associated with heavy jets
at LHC energies and $p_T^2\geq$ 10$^4$ GeV$^2$, when the lifetime of the intrinsic charm is much larger 
than the interaction time, where the intrinsic charm could be resolved.

Taking into account that the IC probability $w$ enters into (\ref{def:cdens_start})
as a constant in front of the function dependent on $x$ and $\mu^2$, one can 
suggest a simple relation at any $w \leq w_{\rm max}$:
\begin{eqnarray}
  xc_{int}(x,\mu^2) = \frac{w}{w_{\rm max}}xc_{int}(x,\mu^2)|_{w = w_{\rm max}}.
\label{def:xcint}
\end{eqnarray}   

\noindent
Actually, that is the linear interpolation between two charm densities at the scale $\mu^2$, 
obtained at $w = w_{\rm max}$ and $w = 0$.
Later we adopt the charm distribution function 
from the CTEQ66M set\cite{27}. We assume $w_{\rm max} = 3.5$\% everywhere, which
corresponds to the CTEQ66c1 set\cite{27}.
Additionally, we performed the three-point interpolation of the charmed quark 
distributions (over $w = 0$, $w = 1$\% and $w = 3.5$\%, which correspond to the 
CTEQ66M, CTEQ66c0 and CTEQ66c1 sets, respectively). These results differ from the ones based on~(\ref{def:xcint}) 
by no more than $0.5$\%, thus giving us the confidence in our starting point. 

Below we apply the charmed quark density obtained by (\ref{def:cdens_start}) and (\ref{def:xcint}) to calculate
the total and differential cross sections of associated prompt photon or $Z$ boson 
and heavy flavor jet production, $\gamma(Z) + Q$, at the LHC conditions.
The suggested procedure to calculate $xc_{int}(x,\mu^2)$ at any $w\leq  w_{\rm max}$ allows us 
to reduce significantly the time for the calculation of these observables.

\section{Theoretical approaches to the associated $\gamma(Z) + Q$ production} \indent

As was mentioned above, we perform the numerical calculations 
of the associated $\gamma(Z) + Q$ production cross sections
using the parton-level Monte Carlo event generator \textsc{mcfm} 
within the NLO pQCD as well as the $k_T$-factorization QCD approach.
The \textsc{mcfm} is able to calculate the processes, that involve the gauge bosons $Z$ or 
$W$ (see\cite{21} for more information).
In contrast to our early study of these processes\cite{13} within the \textsc{mcfm}, we use
this generator to calculate the differential and total cross sections of the $Z + c$ and $Z + b$ production
in the $pp$ collision and their ratio as a function $w$.  

\begin{figure}
\begin{center}
\epsfig{figure=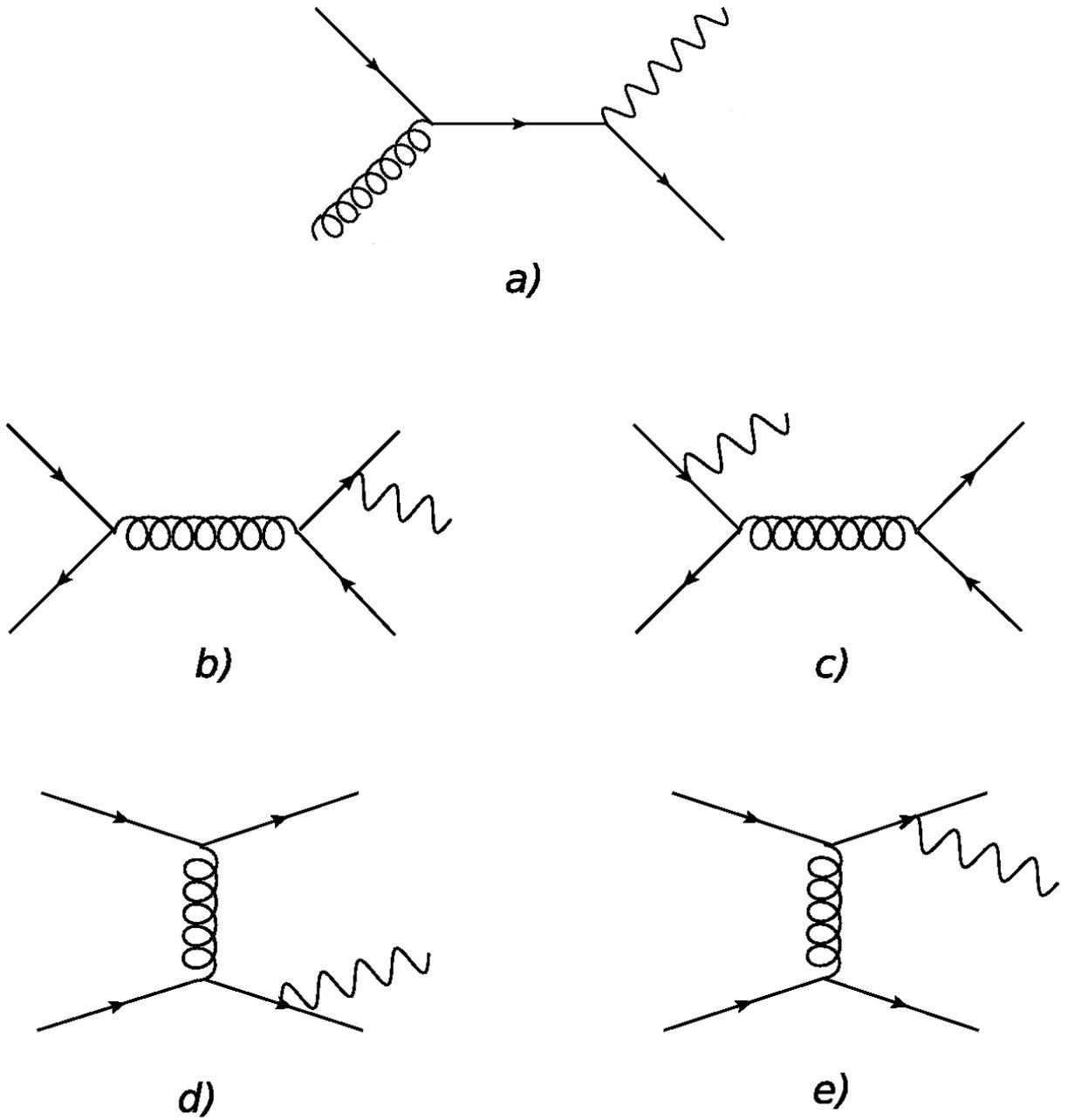, width =16cm}
\vspace{1cm}
\caption{The ${\cal O}(\alpha \alpha_s)$~(a) and ${\cal O}(\alpha \alpha_s^2)$~(b) --- (e) contributions
to the $\gamma(Z) + Q$ production taken into account in the $k_T$-factorization calculations. }
\label{fig2}
\end{center}
\end{figure}

The $k_T$-factorization approach\cite{22,23} is based on the 
small-$x$ Balitsky-Fadin-Kuraev-Lipatov (BFKL)\cite{29} 
gluon dynamics and provides solid theoretical grounds for the effects of the initial gluon 
radiation and the intrinsic parton transverse momentum\footnote{A detailed description of the $k_T$-factorization
approach can be found, for example, in reviews\cite{30}.}.
Our main motivation to use here the $k_T$-factorization formalism
is that its predictions for the associated $\gamma + Q$ production better agree
with the Tevatron data compared to the NLO pQCD (see\cite{17,18}).
The consideration is mainly based on the ${\cal O}(\alpha \alpha_s)$ off-shell 
(depending on the transverse momenta of initial quarks and gluons) 
quark-gluon Compton-like scattering subprocess, see Fig.~\ref{fig2}(a).
Within this approach the transverse momentum dependent (TMD) parton densities
include many high order corrections, while the partonic amplitudes are calculated
within the leading order (LO) of QCD.
The off-shell quark-gluon Compton scattering amplitude is calculated within the  
reggeized parton approach\cite{31,32,33} based on the effective action formalism\cite{34},
which ensures the gauge invariance of the obtained amplitudes 
despite the off-shell initial quarks and gluons\footnote{Here we use the expressions
derived earlier\cite{35}.}.
The TMD parton densities are calculated using the Kimber-Martin-Ryskin (KMR)
approach, currently developed within the NLO\cite{36}.
This approach is the formalism to construct the TMD 
quark and gluon densities from the known conventional parton distributions. 
The key assumption is that the $k_T$ dependence appears 
at the last evolution step, so that the 
DGLAP evolution 
can be used up to this step. 
Numerically, for the input we used parton densities derived in Section~2.
Other details of these calculations are explained in\cite{35}.

To improve the $k_T$-factorization predictions at high transverse momenta, we 
take into account some ${\cal O}(\alpha \alpha_s^2)$ contributions,
namely 
$q \bar q \to V Q{\bar Q}$ and $q Q \to V q Q $ ones, where 
$V$ denotes the photon or the $Z$ boson, see Fig.~\ref{fig2}(b) --- (e).
These contributions are significant at large $x$ and therefore
can be calculated in the usual collinear QCD factorization scheme.
Thus, we rely on the combination of two techniques that is most suitable. 

\section{Results and discussion} \indent

Let us present the results of our calculations.  First of all we describe our numerical input.
Following to\cite{37}, we set the charmed and beauty quark masses
$m_c = 1.4$~GeV, $m_b = 4.75$~GeV, the $Z$-boson mass 
$m_Z = 91.1876$~GeV, and $\sin^2 \theta_W = 0.23122$. 
The chosen factorization and renormalization scales are 
$\mu_R = \mu_F = \xi p_T$ or $\mu_R = \mu_F = \xi m_T$, 
where $p_T$ is the produced photon 
transverse momentum and $m_T$ is the $Z$ boson transverse mass. 
As usual, we vary the nonphysical parameter $\xi$ between $1/2$ and $2$ about the default value
$\xi = 1$ in order to estimate the scale uncertainties of our calculations.
We employ the two-loop formula for the strong coupling constant
with active quark flavors $n_f = 5$ at $\Lambda_{\rm QCD} = 226.2$~MeV
and use the running QED coupling constant over a 
wide region of transverse momenta.
The multidimensional integration in the $k_T$-factorization calculations
was performed by means of the Monte Carlo technique, 
using the \textsc{vegas} routine\cite{38}.

In our calculations we also follow the conclusion obtained in our papers\cite{9,13} that the IC
signal in the hard processes discussed here can be detected at ATLAS or CMS of the LHC in the forward rapidity
region $1.5 < |\eta|< 2.4$ and  $p_T > 50$~GeV.   
Additionally, we require $|\eta(Q)| < 2.4$ and $p_T(Q) > 25$~GeV,
where $\eta(Q)$ and $p_T(Q)$ are the pseudo-rapidity and transverse 
momentum of the heavy quark jet in a final state, as was done in\cite{9,13}.

\begin{figure}
\begin{center}
\epsfig{figure=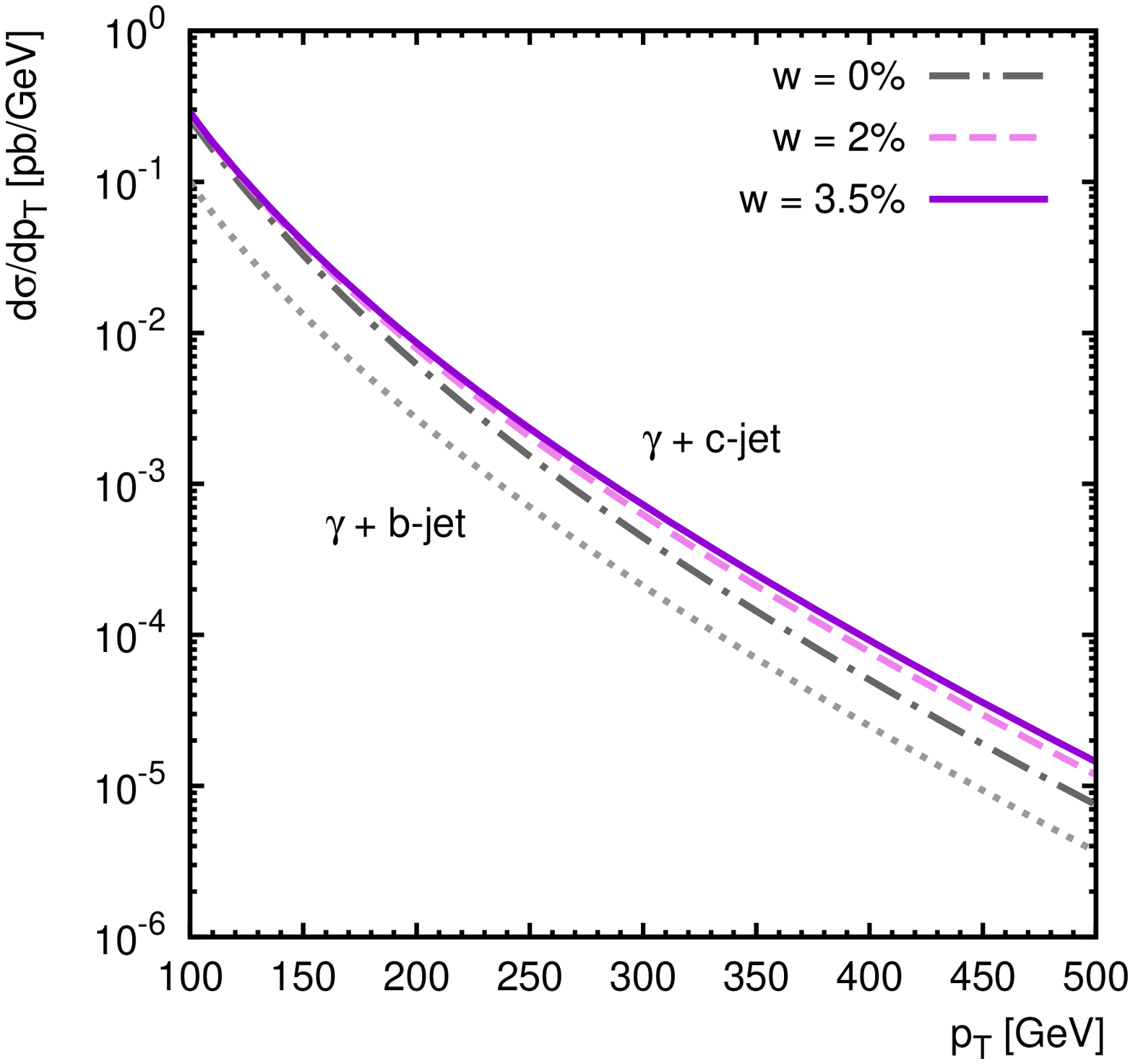,width = 10cm, angle = 270}
\hspace{-1.0cm}
\epsfig{figure=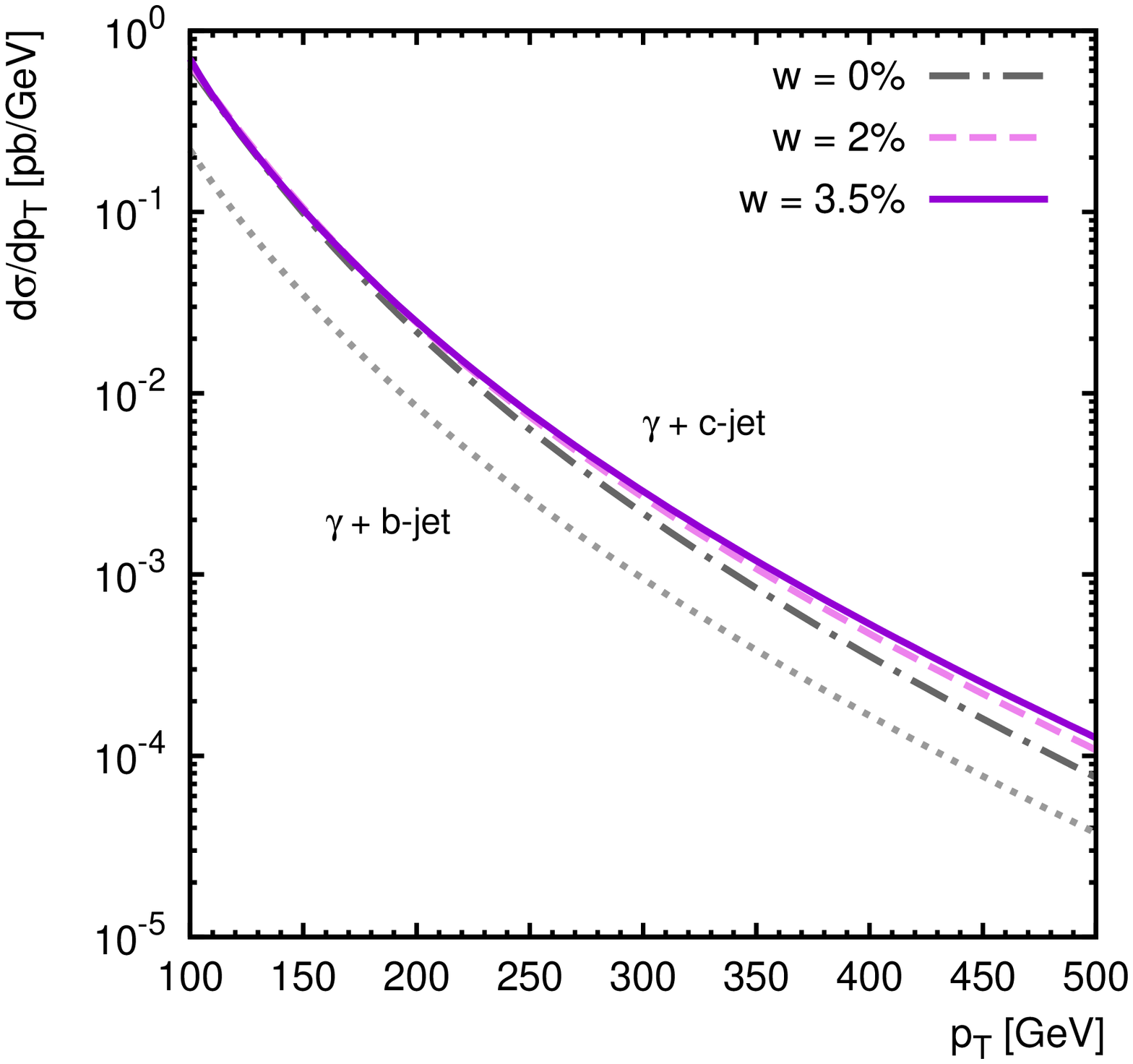,width = 10cm,angle = 270}
\caption{The cross sections of the associated $\gamma + c$ and $\gamma + b$ production in the $pp$ collision 
calculated as a function of the photon transverse momentum $p_T$
at $\sqrt s = 8$~TeV (top) and $\sqrt s = 13$~TeV (bottom) within the $k_T$-factorization approach.
The kinematical conditions are described in the text.}
\label{fig3}
\end{center}
\end{figure}

\begin{figure}
\begin{center}
\epsfig{figure=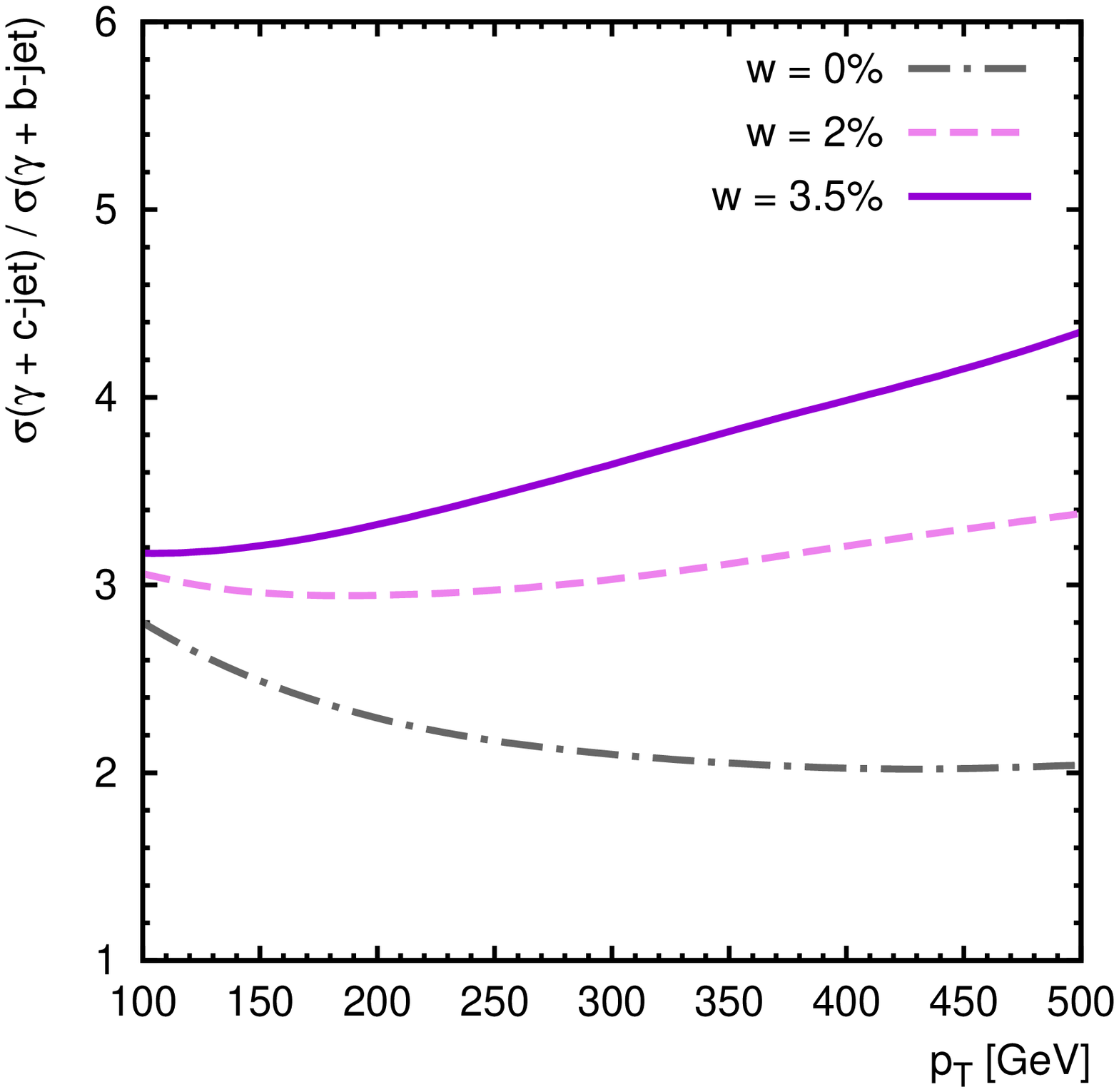,width = 10cm, angle = 270}
\hspace{-1.0cm}
\epsfig{figure=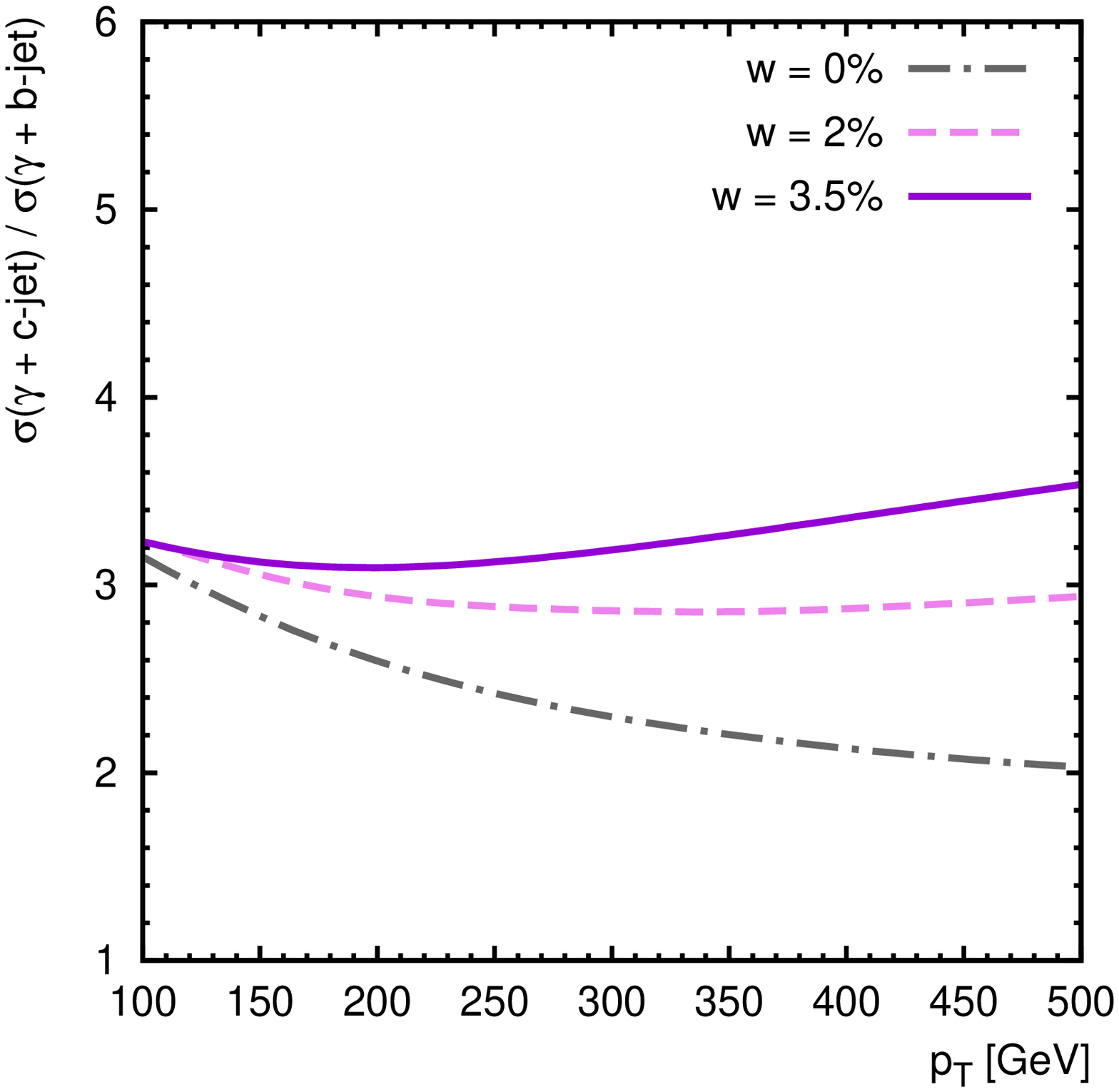,width = 10cm,angle = 270}
\caption{The cross section ratio of the $\gamma + c$ production to the $\gamma + b$ one in the $pp$ collision 
calculated as a function of the photon transverse momentum $p_T$
at $\sqrt s = 8$~TeV (top) and $\sqrt s = 13$~TeV (bottom) within the $k_T$-factorization approach.
The kinematical conditions are described in the text.}
\label{fig3_grat}
\end{center}
\end{figure}

\begin{figure}
\begin{center}
\epsfig{figure=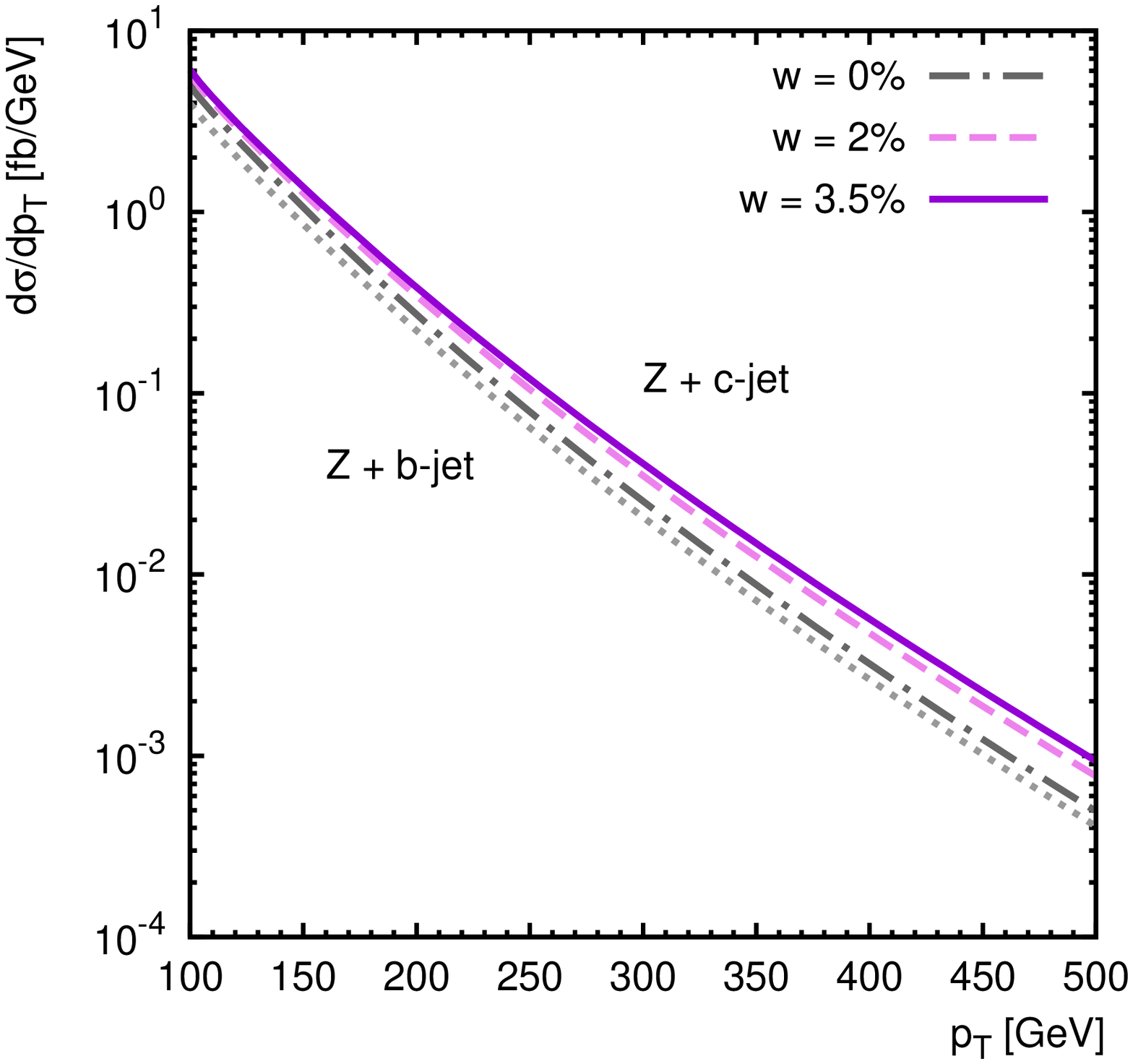,width = 10cm, angle = 270}
\hspace{-1.0cm}
\epsfig{figure=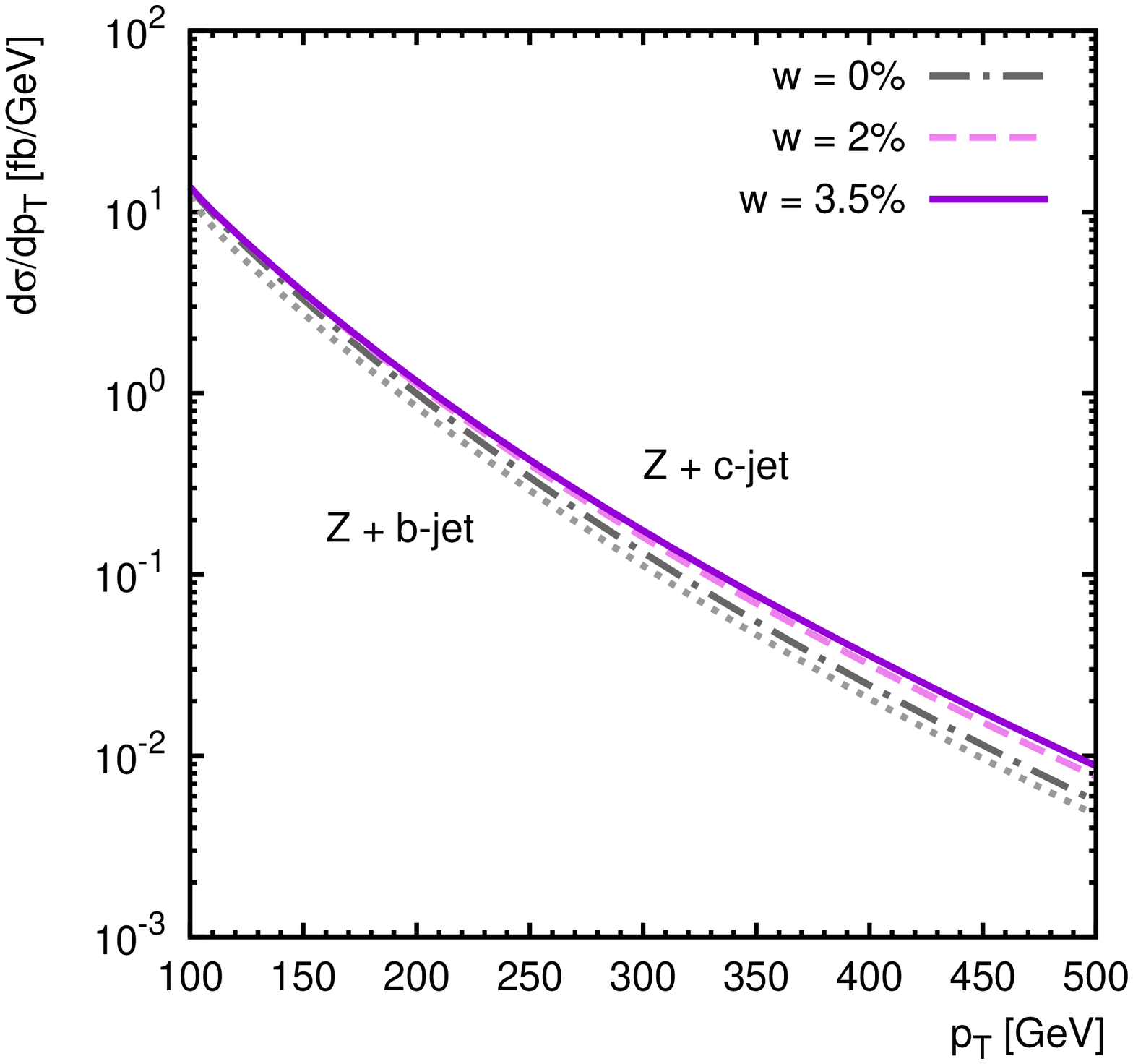,width = 10cm,angle = 270}
\caption{The cross sections of the associated $Z + c$ and $Z + b$ production in the $pp$ collision 
calculated as a function of the $Z$ boson transverse momentum $p_T$
at $\sqrt s = 8$~TeV (top) and $\sqrt s = 13$~TeV (bottom) within the $k_T$-factorization approach.
The kinematical conditions are described in the text.}
\label{fig4}
\end{center}
\end{figure}

\begin{figure}
\begin{center}
\epsfig{figure=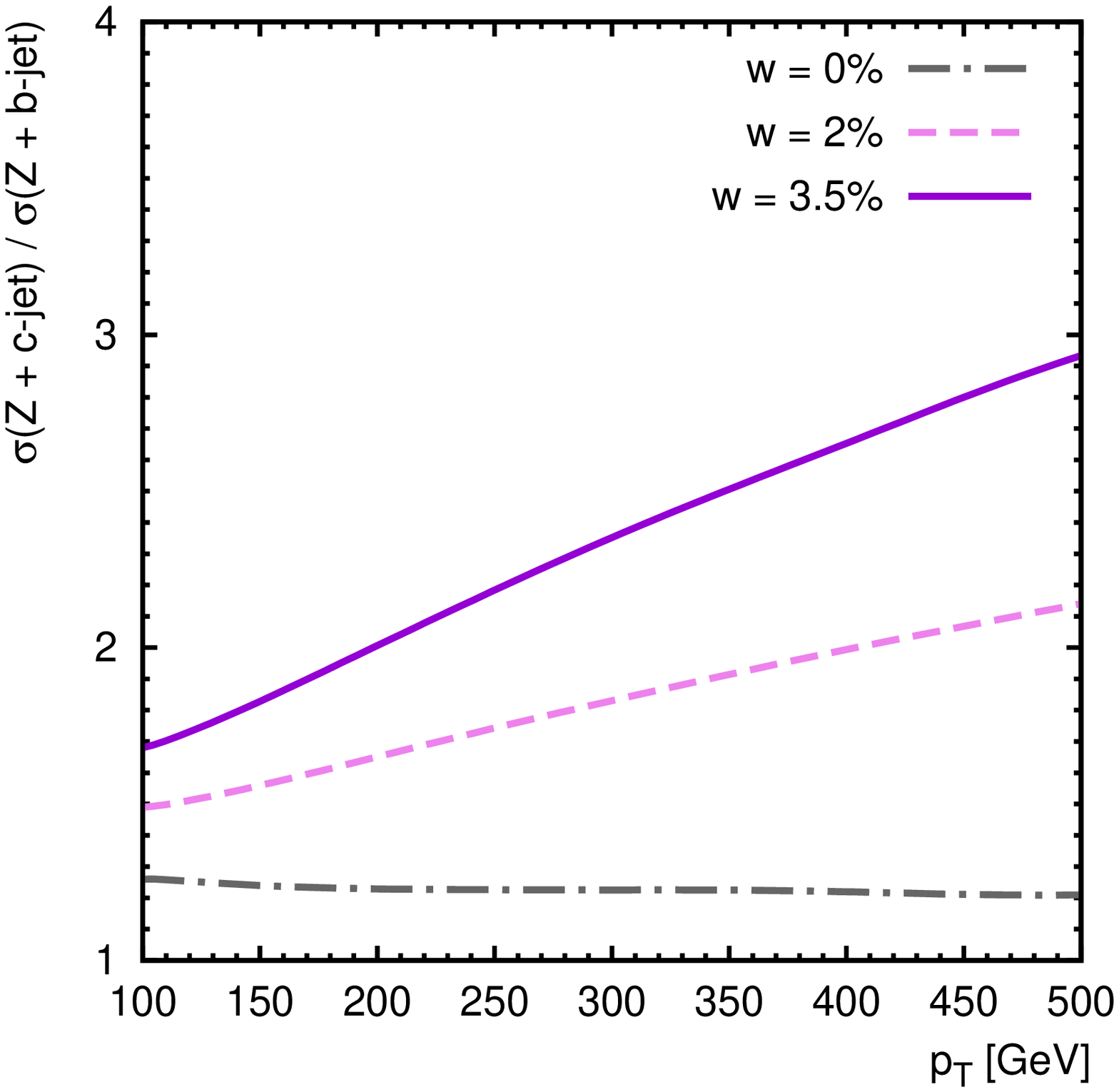,width = 10cm, angle = 270}
\hspace{-1.0cm}
\epsfig{figure=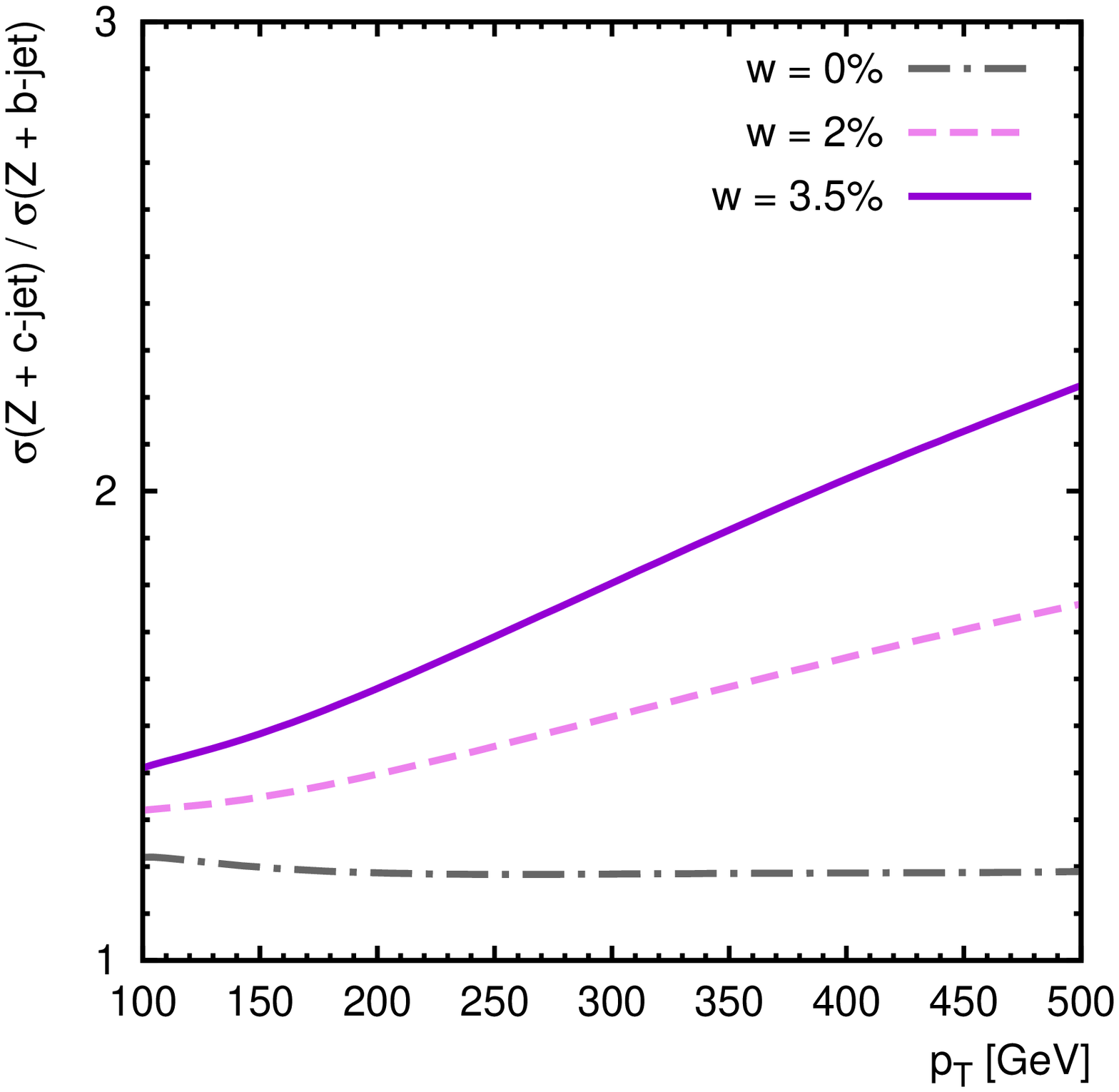,width = 10cm,angle = 270}
\caption{The cross section ratio of the $Z + c$ production to the $Z + b$ one in the $pp$ collision 
calculated as a function of the $Z$ boson transverse momentum $p_T$
at $\sqrt s = 8$~TeV (top) and $\sqrt s = 13$~TeV (bottom) within the $k_T$-factorization approach.
The kinematical conditions are described in the text.}
\label{fig4_Zrat}
\end{center}
\end{figure}

\begin{figure}
\begin{center}
\epsfig{figure=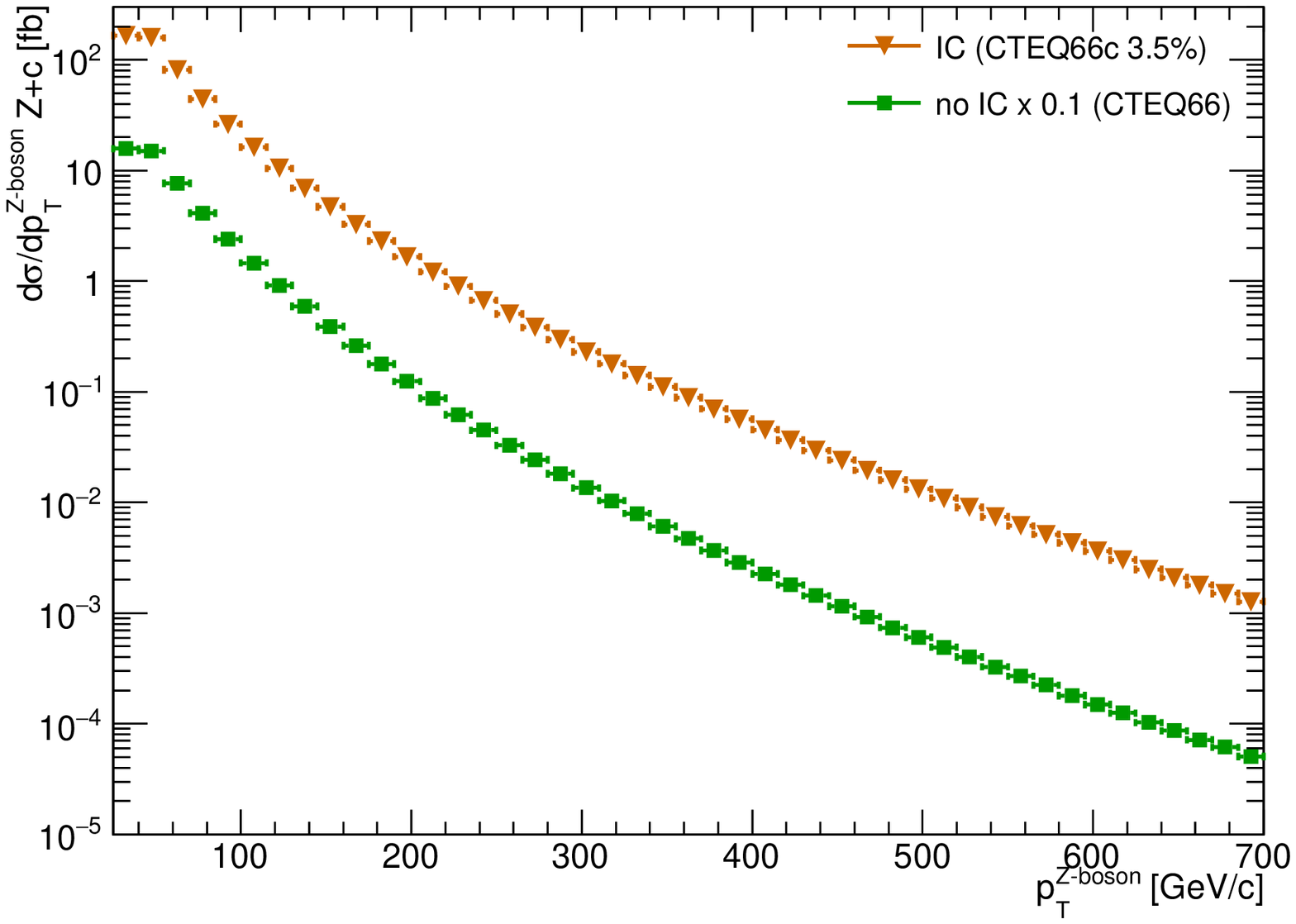, width = 10cm}
\epsfig{figure=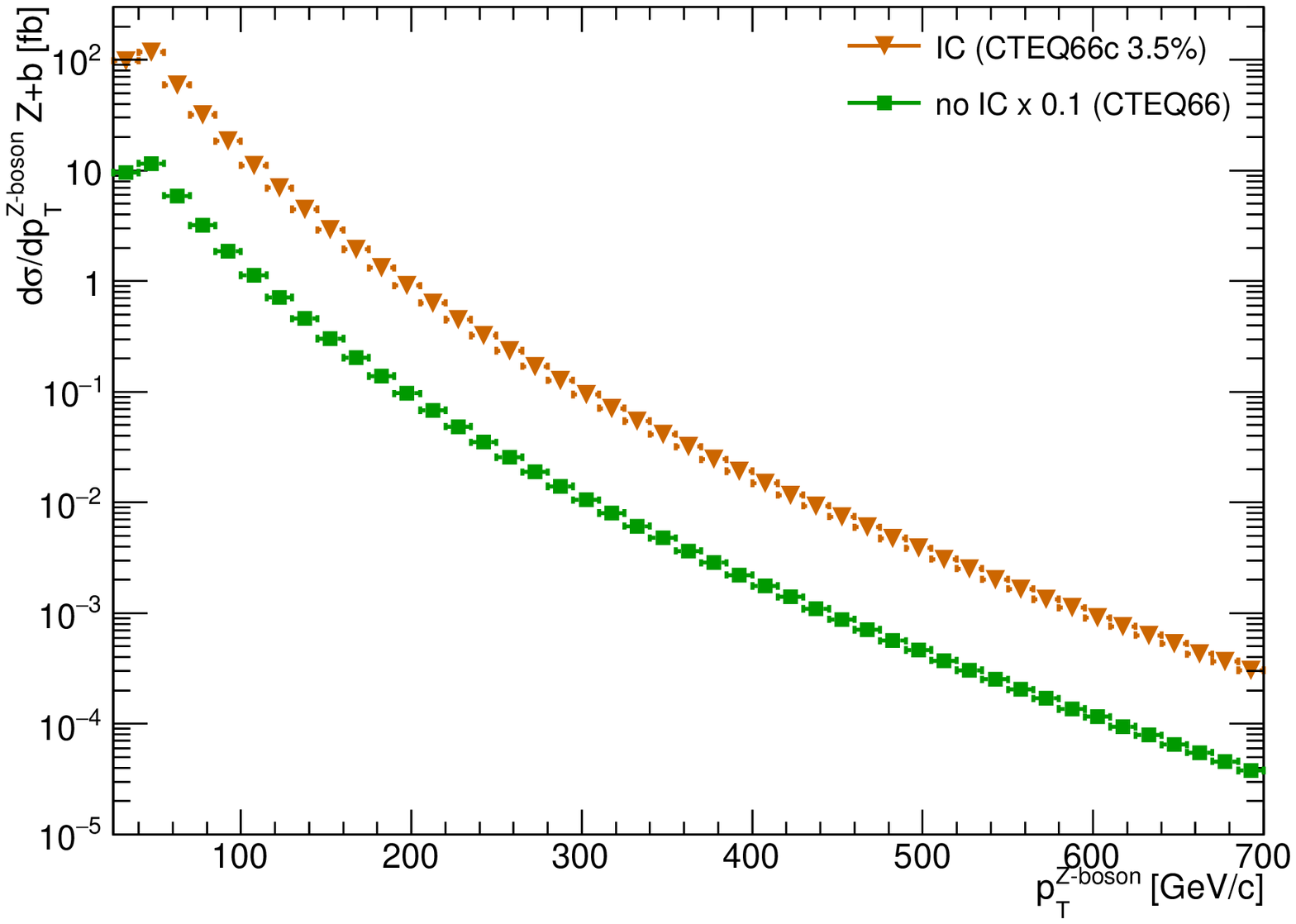, width = 10cm}
\caption{The cross sections of the associated $Z + c$ (top) and $Z + b$ (bottom) 
production in $pp$ collision calculated as a function of the $Z$ boson transverse momentum $p_T$
at $\sqrt s = 13$~TeV within the \textsc{mcfm} routine.
The kinematical conditions are described in the text.}
\label{fig5}
\end{center}
\end{figure}

\begin{figure}
\begin{center}
\epsfig{figure=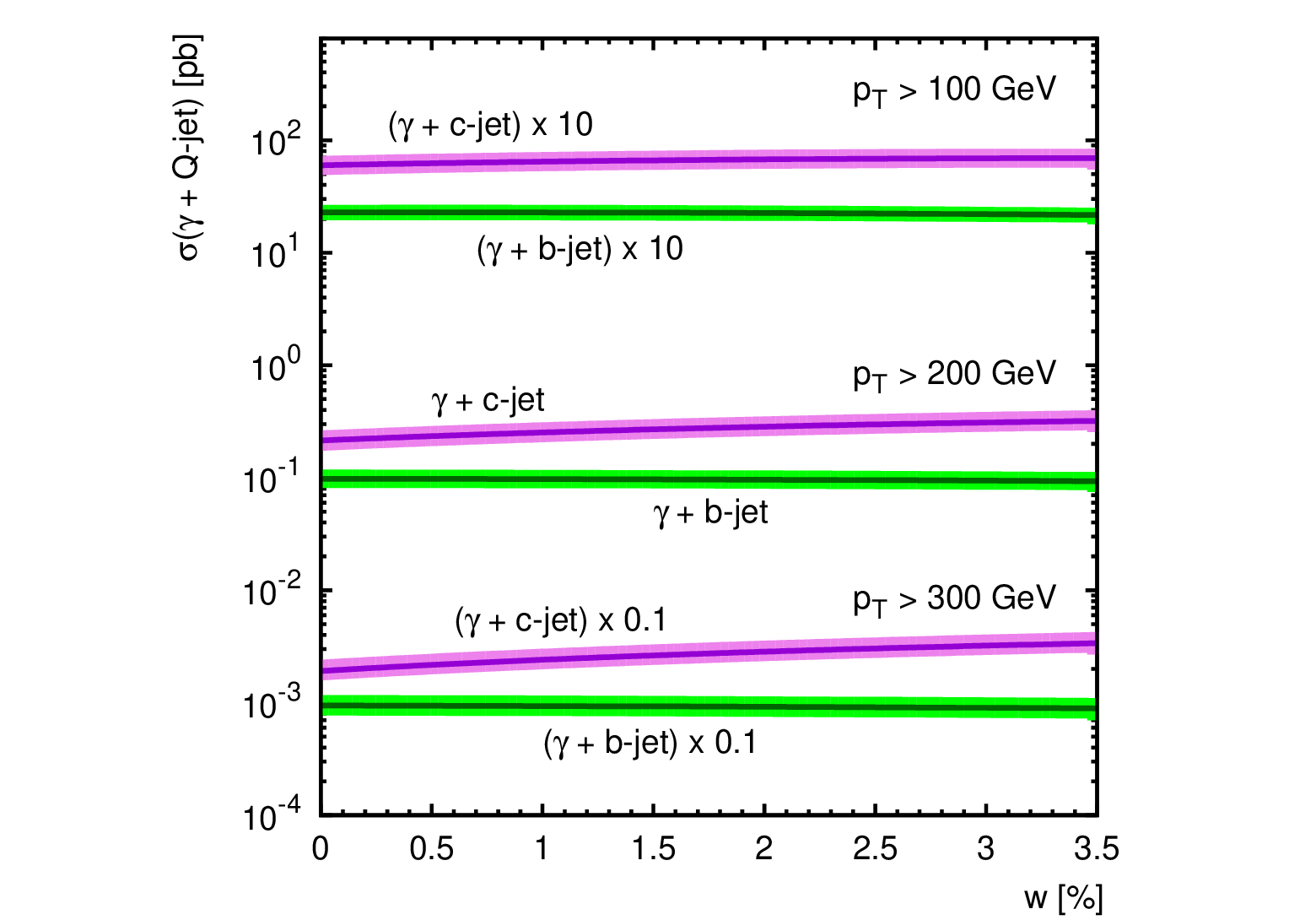, width = 5.8cm, angle = 270} 
\vspace{0.7cm} \hspace{-1cm}
\epsfig{figure=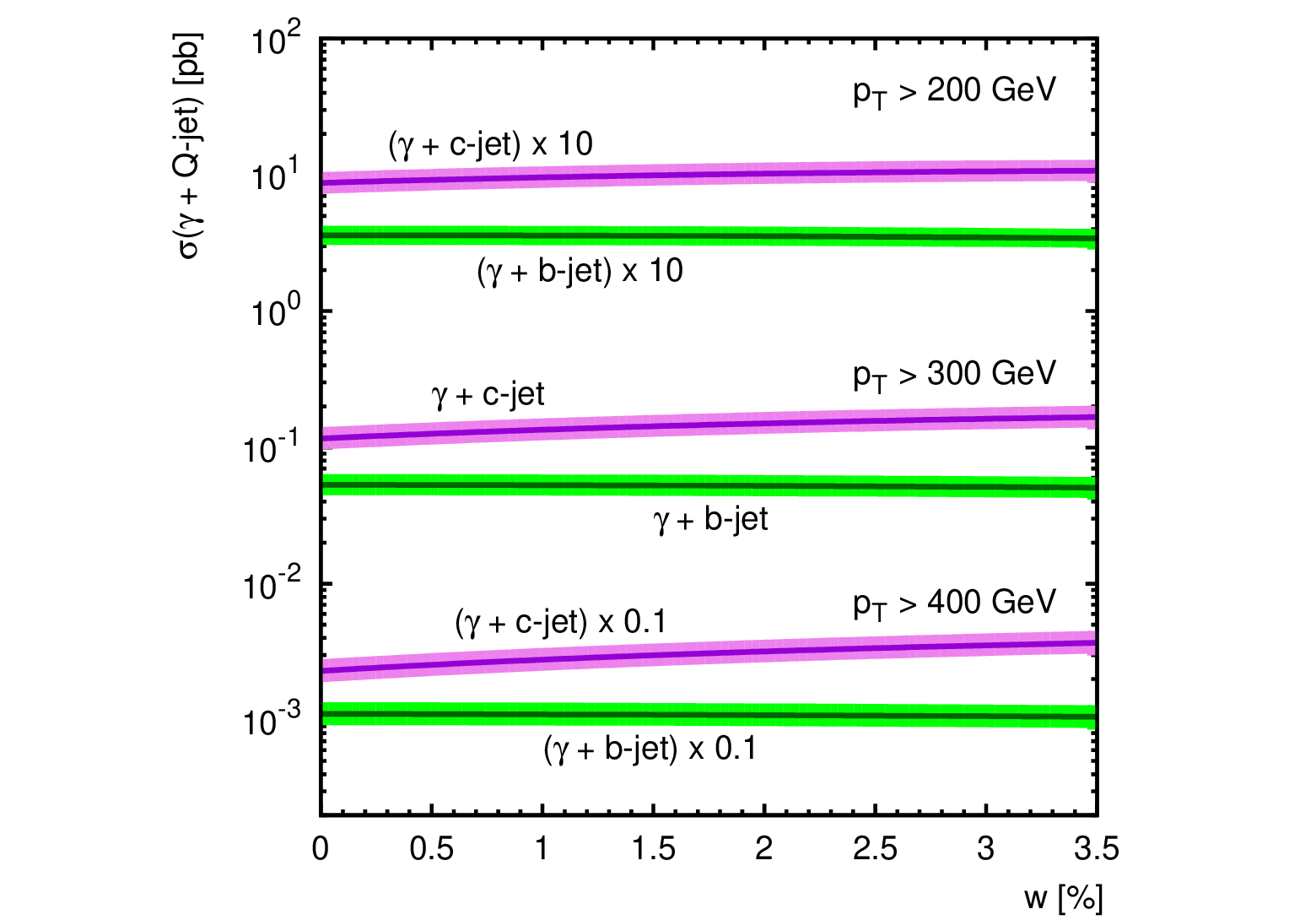, width = 5.8cm, angle = 270}
\vspace{0.7cm}
\epsfig{figure=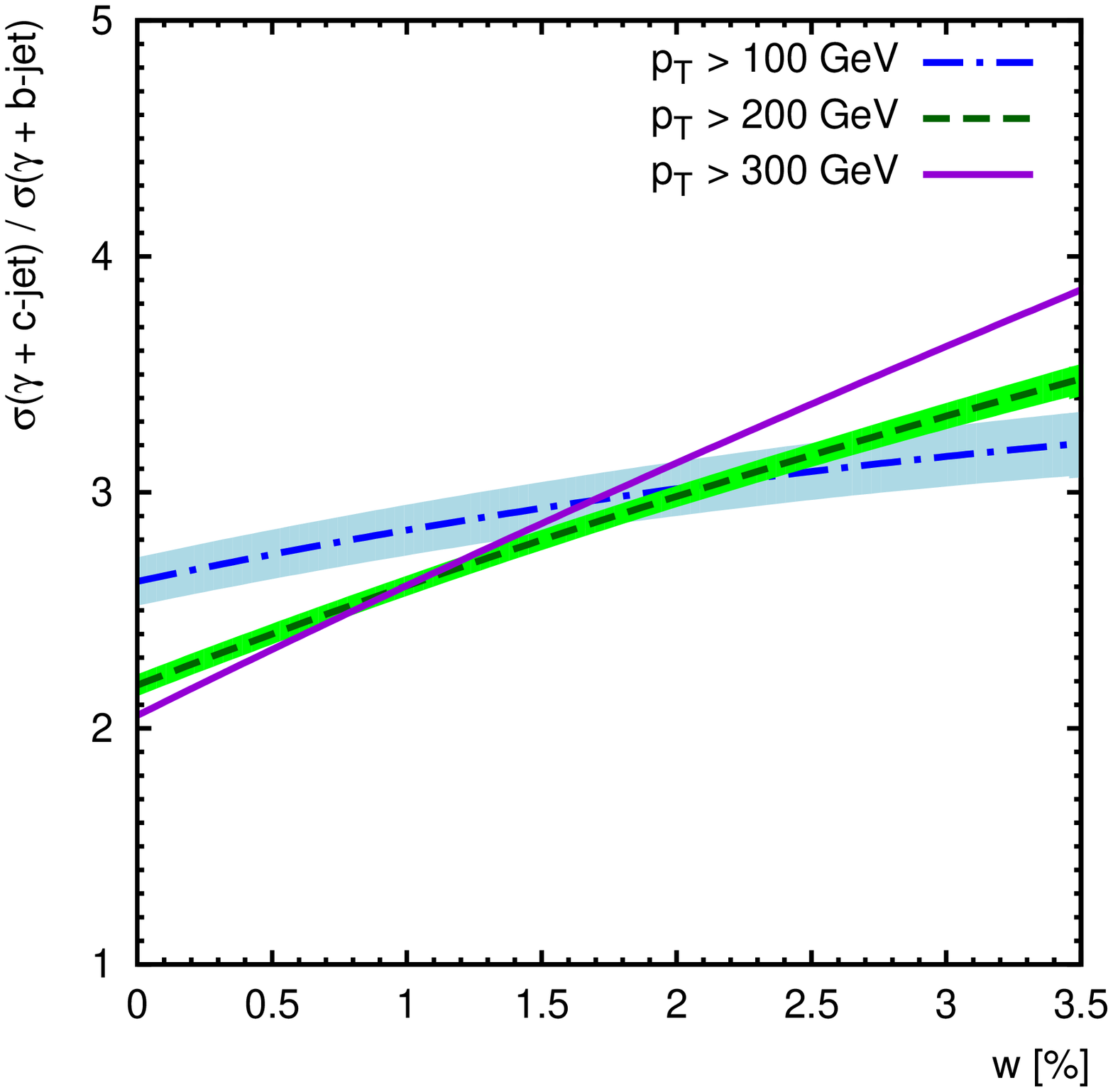 , width = 5.8cm, angle = 270}
\hspace{-1cm}
\epsfig{figure=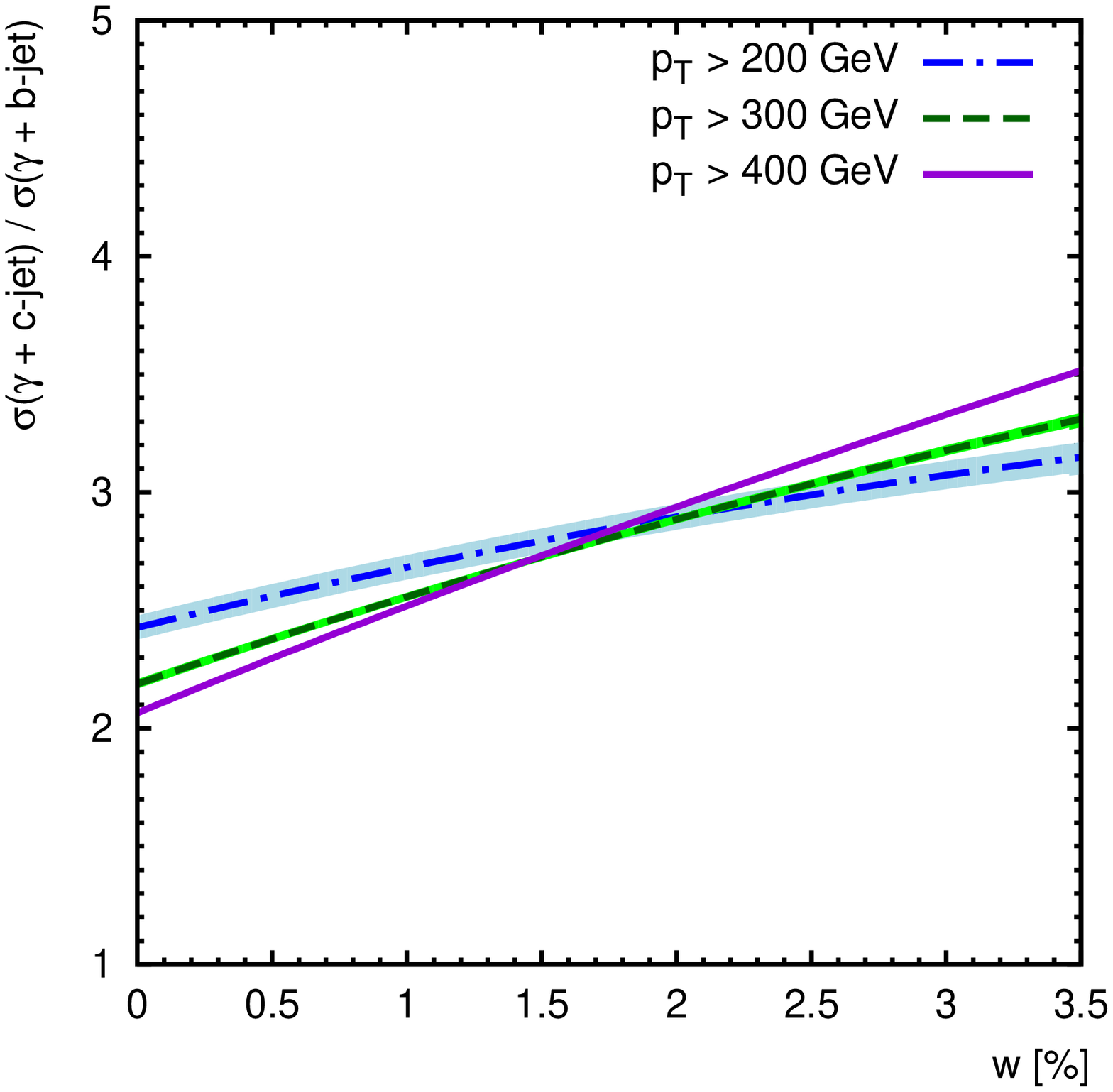, width = 5.8cm, angle = 270}
\caption{Top: the cross sections of the associated $\gamma + c$ and $\gamma + b$ 
production in the $pp$ collision as a function of $w$ integrated over the photon 
transverse momenta $p_T > p_T^{\rm min}$ for 
different $p_T^{\rm min}$ at $\sqrt{s} = 8$~TeV (left)
and $\sqrt{s} = 13$~TeV (right). The kinematical conditions are described in the text.
Bottom: the corresponding ratios
of these cross sections.
The calculations were done using the $k_T$-factorization approach.
The bands correspond to the usual scale variation as it is described in the text.}
\label{fig7}
\end{center}
\end{figure}

\begin{figure}
\begin{center}
\epsfig{figure=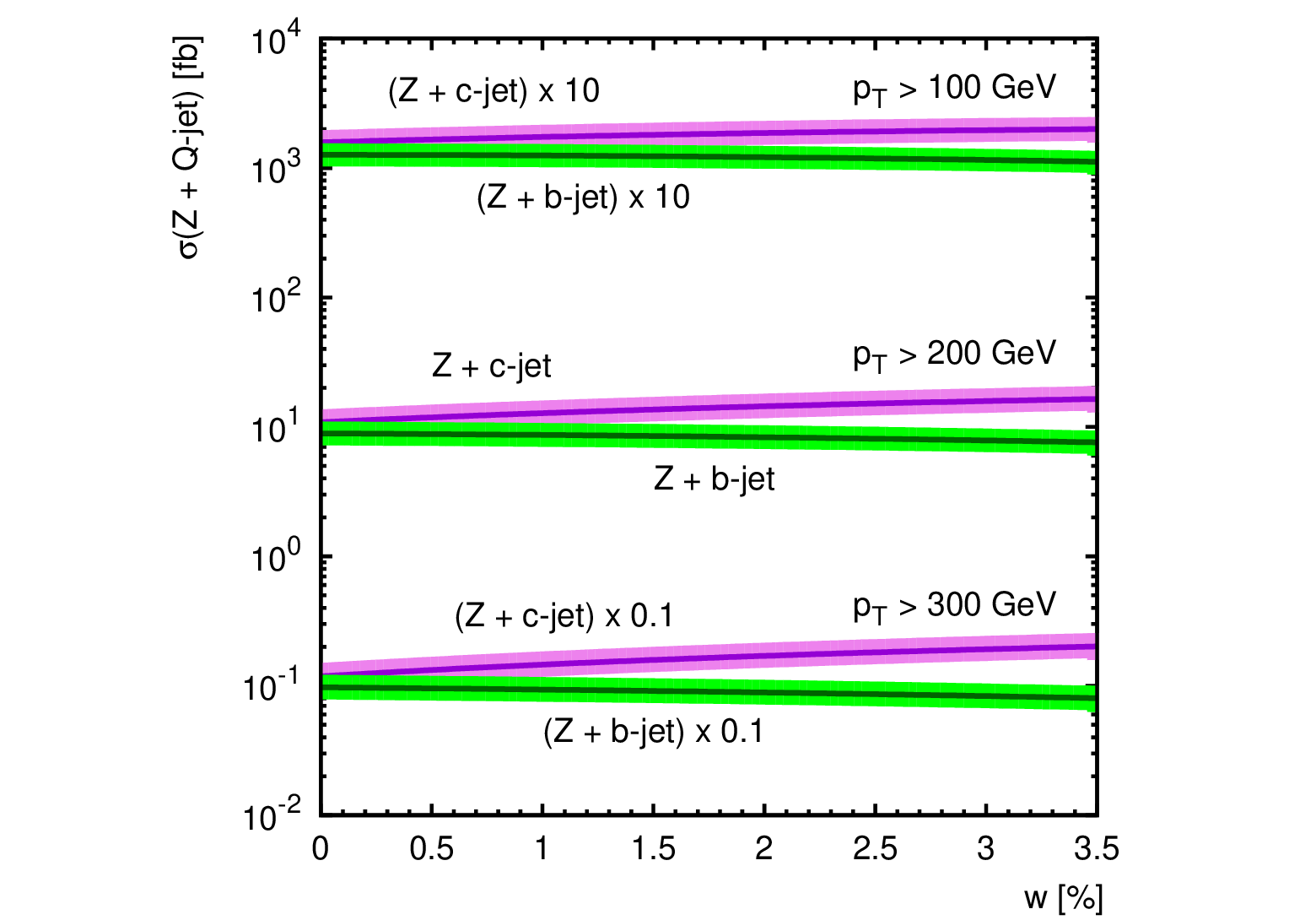, width = 5.8cm, angle = 270}
\vspace{0.7cm} \hspace{-1cm}
\epsfig{figure=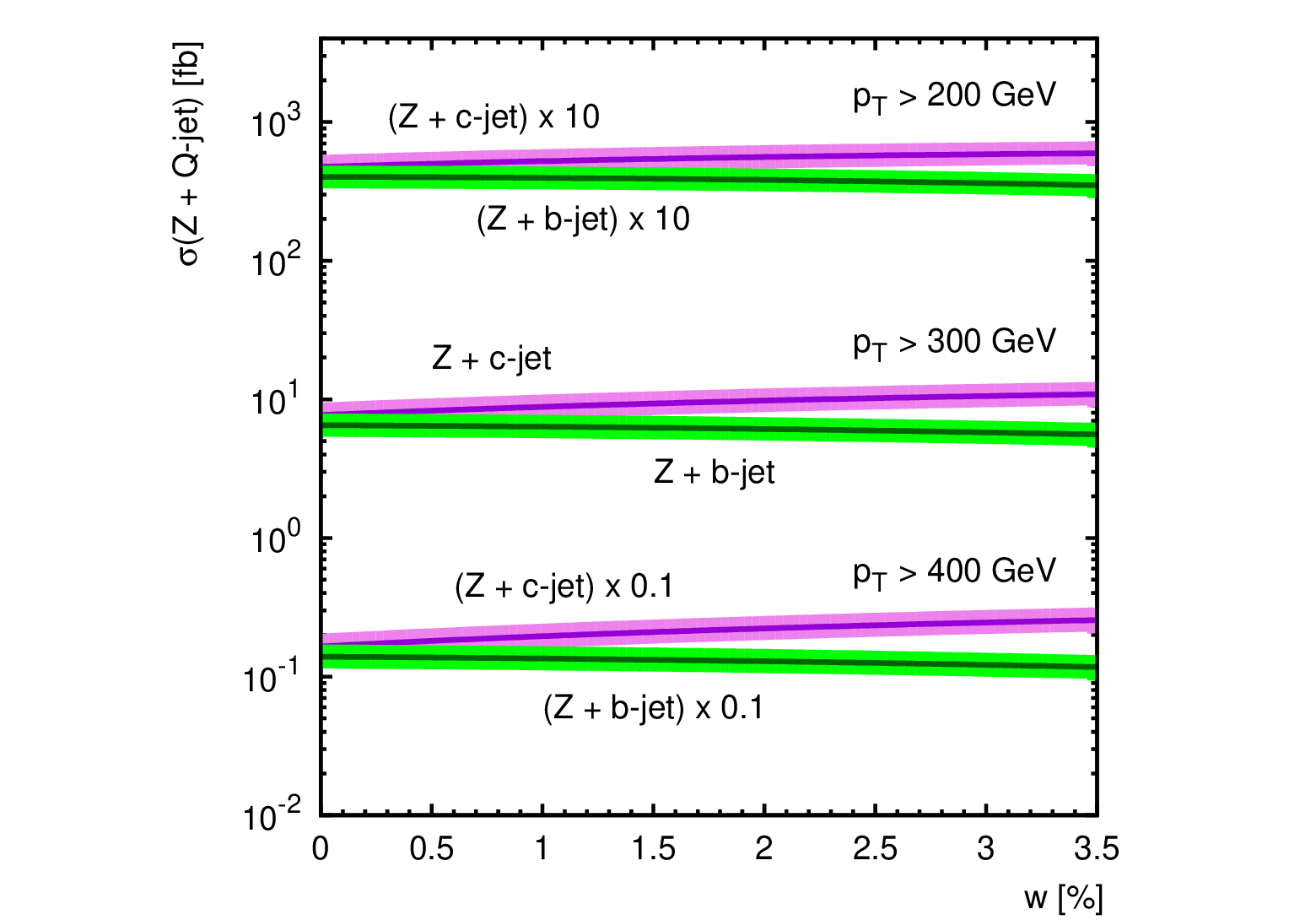, width = 5.8cm, angle = 270}
\vspace{0.7cm}
\epsfig{figure=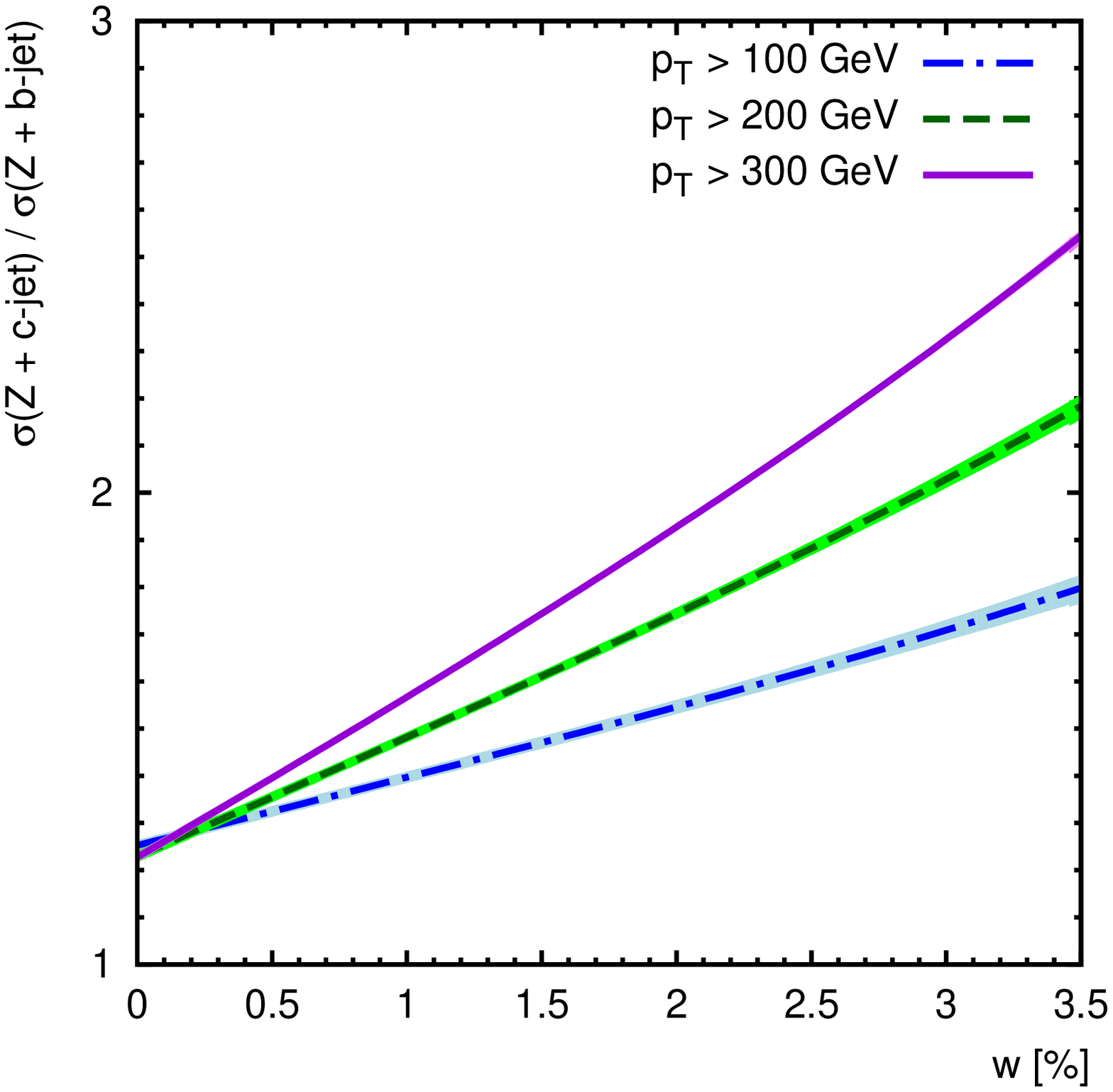, width = 5.8cm, angle = 270}
\hspace{-1cm}
\epsfig{figure=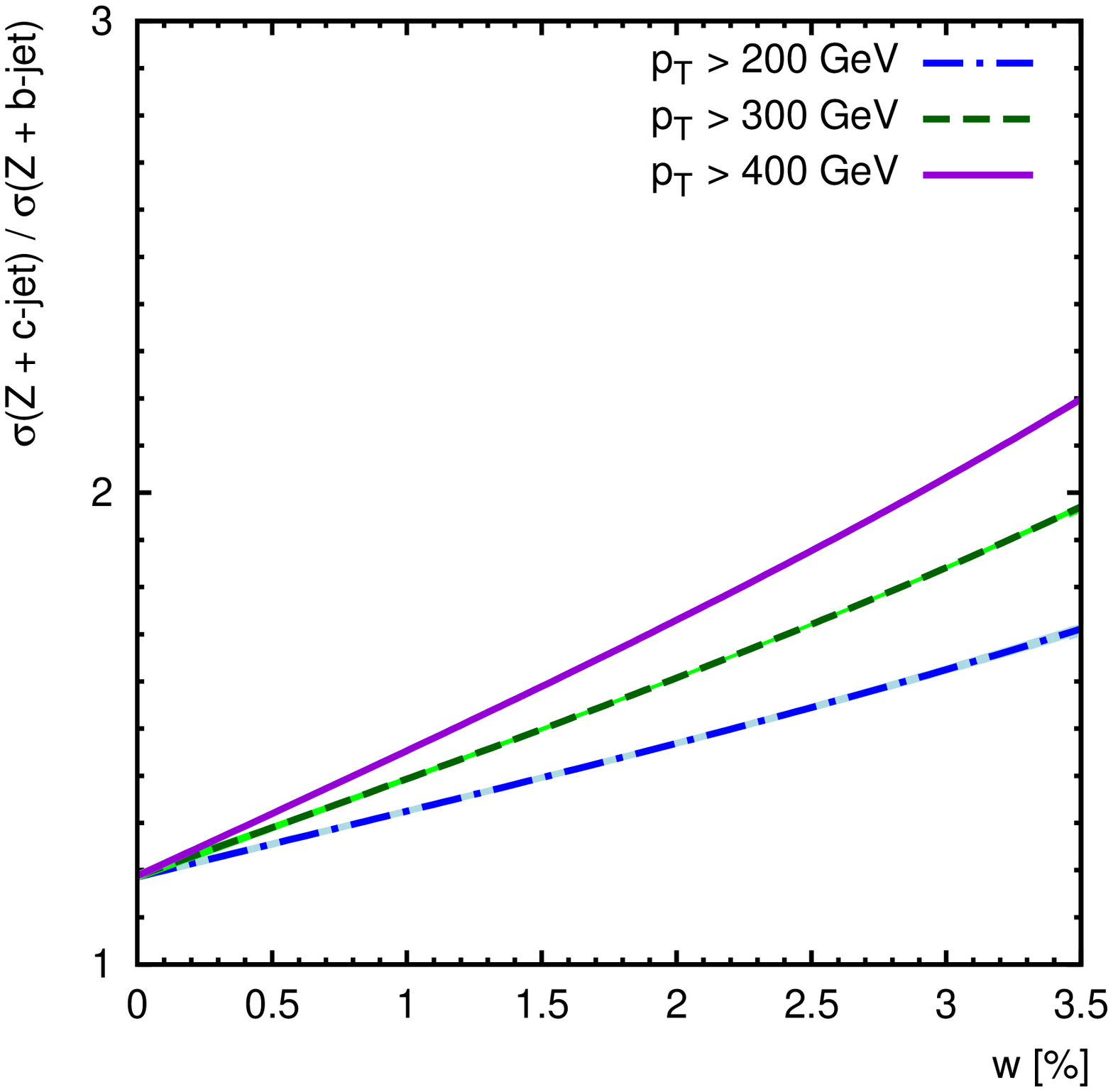, width = 5.8cm, angle = 270}

\caption{Top: the cross sections of the associated $Z + c$ and $Z + b$ 
production in the $pp$ collision as a function of $w$ integrated over the $Z$ boson 
transverse momenta $p_T > p_T^{\rm min}$ for 
different $p_T^{\rm min}$ at $\sqrt{s} = 8$~TeV (left
and $\sqrt{s} = 13$~TeV (right). The kinematical conditions are described in the text. 
Bottom: the corresponding ratios of these cross sections.
The calculations were done using the $k_T$-factorization approach.
The bands correspond to the usual scale variation as it is described in the text.} 
\label{fig8}
\end{center}
\end{figure}

The results of our calculations are shown in Figs.~\ref{fig3} --- \ref{fig8}.
The transverse momentum distributions of photons and $Z$ bosons accompanied by the $c$ and $b$
quarks are presented in Figs.~\ref{fig3}, \ref{fig4} and \ref{fig5}
at the different IC probability $w$ (namely, $w = 0$\%, $w = 2 $\% and $w = 3.5$\%)
at $\sqrt{s} = 8$ and $13$~TeV.
One can see in Figs.~\ref{fig4} and \ref{fig5} that the \textsc{mcfm} and $k_T$-factorization predictions
for $Z + Q$ production are very similar in the whole $p_T$ region, therefore below we 
will present the observables calculated within the $k_T$-factorization approach only.
The coincidence of these two calculations is due to effective allowance for the high-order corrections 
within the $k_T$-factorization formalism (see, for example,\cite{30} for more information).
Both types of calculations predict a significant enhancement of
$p_T$ distributions due to the IC terms at $p_T \geq 100$~GeV,
which is in agreement with the previous studies\cite{9,12,13}.
 
The $p_T$-spectrum ratios $\sigma(\gamma+c)/\sigma(\gamma+b)$ and $\sigma(Z+c)/\sigma(Z+b)$ 
versus $p_T$ at different 
$w$ are presented in Figs.~\ref{fig3_grat} and~\ref{fig4_Zrat}.
One can see that in the absence of the IC contribution 
the ratio $\sigma(\gamma+c)/\sigma(\gamma+b)$ is about $3$ at
$p_T \sim 100$~GeV and decreases down to $2$ at $p_T \sim 500$~GeV.
This behavior is the same for both energies $\sqrt{s} = 8$~TeV and $\sqrt{s} = 13$~TeV.
If one takes into account the IC contributions, this ratio
becomes approximately flat at $w = 2$\% or even increasing up to about $4$ at $w = 3.5$\%.  
It is very close to the Tevatron data\cite{20}:
the constant ratio $\sigma(\gamma + c) / \sigma (\gamma + b) \sim 3.5 - 4.5$
measured in the $p\bar p$ collisions at $110 < p_T < 300$~GeV and $\sqrt s = 1.96$~TeV.
However, this agreement cannot be treated as the IC indication
due to huge experimental uncertainties (about $50$\%) and rather different kinematical 
conditions.
If the IC contribution is included, the ratio $\sigma(Z+c)/\sigma(Z+b)$ also increases by a factor about $2$ at $w = 3.5$\%,
when the $Z$ boson transverse momentum grows from $100$~GeV to $500$~GeV (see Fig.~\ref{fig4_Zrat}).
In the absence of the IC terms this ratio slowly decreases.

One can consider other observables which could be useful
to detect the IC signal, the  
cross sections discussed above but integrated over $p_T > p_T^{\rm min}$, where 
$p_T^{\rm min} \geq 100$~GeV, and their ratios. 
Our predictions for such integrated cross sections versus the IC probability $w$
at $p_T^{\rm min} = 100$, $200$ and $300$~GeV for $\sqrt s = 8$~TeV and 
$p_T^{\rm min} = 200$, $300$ and $400$~GeV for $\sqrt s = 13$~TeV are shown 
in Figs.~\ref{fig7} and \ref{fig8}. 

All the $p_T$-spectra have a significant scale uncertainty as is shown in\cite{13} (see also Figs.~\ref{fig7} and \ref{fig8}). 
According to\cite{13}, the ratio between the cross sections for the $Z+Q$ 
and $W+Q$ production in the $pp$ collision is less sensitive to the scale variation calculated within the \textsc{mcfm}.
Nevertheless, the uncertainty in this ratio at large $p_T > 250$~GeV is about 40 --- 50$\%$.
In the present paper we check these results for the ratios 
$\sigma(\gamma+c)/\sigma(\gamma+b)$ and $\sigma(Z+c)/\sigma(Z+b)$. 
In Figs.~\ref{fig7} and~\ref{fig8} (bottom) we present these ratios 
versus the IC probability $w$ calculated
at different scales, 
when the cross sections of $\gamma(Z) + Q$ production are integrated within 
the different intervals of transverse momentum. One can see a very small QCD scale uncertainty, especially at 
$\sqrt{s} = 13$~TeV (bottom right), which is less than 1$\%$. In contrast, the scale uncertainty for the 
integrated $\gamma(Z) + Q$ cross sections (see Figs.~\ref{fig7} and~\ref{fig8}, top) 
is significant and amounts to about $30$ --- $40$\%. 
The sizable difference between the scale uncertainties for the ratios 
$\sigma(Z+Q)/\sigma(W+Q)$ and $\sigma(Z+c)/\sigma(Z+b)$ is due to the different matrix elements 
for the $Z+Q$ and $W+Q $ production in $pp$ collisions, while the 
matrix elements for the $Z+c$ and $Z+b$ production are the same.

It is important that the calculated ratios
$\sigma(\gamma+c)/\sigma(\gamma+b)$ and $\sigma(Z+c)/\sigma(Z+b)$
can be used to determine the IC probability $w$ from the future LHC data. 
Moreover, these ratios are practically independent of the uncertainties of our 
calculations: actually, the curves corresponding to the usual scale variations as described above
coincide with each other (see Figs.~\ref{fig7} and~\ref{fig8}, bottom).
Therefore, we can recommend these observables as 
a test for the hypothesis of the IC component inside the proton.


\section{Conclusion} \indent
 
The transverse momentum spectra of the prompt photons and $Z$ bosons produced in association with the $c$ or $b$ jets in 
$pp$ collisions are calculated using the \textsc{mcfm} (NLO pQCD) and the $k_T$-factorization approach at the 
LHC energies and pseudo-rapidites $1.5 < \eta < 2.4$ using PDFs with and without the IC contribution. 
It is shown that these two approaches give similar results. 
We found that the contribution of the intrinsic charm can give a significant signal in the ratios  
$\sigma(\gamma + c) / \sigma(\gamma + b)$ and 
$\sigma(Z + c) / \sigma(Z + b)$ at forward pseudo-rapidities ($1.5 < \eta <2.4$) corresponding 
to the ATLAS and CMS facilities.
If the IC contributions are taken into account,
the ratio $\sigma(\gamma + c) / \sigma(\gamma + b)$ as a function of 
the photon transverse momentum is approximately flat or increases at 
$p_T > 100$~GeV. The similar flat behavior of this ratio was observed in the $p{\bar p}$ annihilation at the Tevatron.
In the absence of the IC contributions this ratio decreases.
Similarly, the ratio $\sigma(Z + c) / \sigma (Z + b)$ increases when the $Z$ boson 
transverse momentum grows if the IC contribution is included and
slowly decreases in the absence of the IC terms.
We argued that the ratio of
the cross sections 
$\gamma(Z) + c$ and $\gamma(Z) + b$
integrated over $p_T > p_T^{\rm min}$ with $p_T^{\rm min} \geq 100$~GeV
can be used to determine the IC probability from the future LHC data. 
The advantage of the proposed ratios is that the theoretical uncertainties
are very small, while the uncertainties for the $p_T$-spectra of photons or $Z$ bosons produced 
in association with the $c$ or $b$ jets are large. Therefore, the search for the IC signal by
analyzing the ratio $\sigma(\gamma/Z + c) / \sigma(\gamma/Z + b)$ can be more promising.

\section{Acknowledgments} \indent

We thank S.J.~Brodsky, A.A.~Glasov and D.~Stump for extremely helpful discussions and 
recommendations in the study of this topic. 
The authors are grateful to H.~Jung, P.M.~Nadolsky 
for very useful discussions and comments. The authors are also grateful to 
L.~Rotali for very constructive discussions.  
This work was supported in part by grant of the President of Russian Federation NS-7989.2016.2.
A.V.L. is grateful to the DESY Directorate for the
support within the framework of the Moscow --- DESY project on Monte-Carlo 
implementation for HERA --- LHC.

\newpage

\end{document}